\documentclass[pra,
twocolumn,a4paper,superscriptaddress,floatfix]{revtex4}

\usepackage{graphicx}
\usepackage{amsmath}
\usepackage{bm}
\usepackage{amssymb}
\usepackage{color}
\usepackage[normalem]{ulem}

\newcommand{\vect}[1]{{\mathbf #1}}

\newcommand{\eF}{\varepsilon_F}
\newcommand{\eB}{\varepsilon_B}

\begin{document}

\title{BCS-BEC crossover in a quasi-two-dimensional Fermi gas}

\author{Andrea M.\ Fischer}
\affiliation{Cavendish Laboratory, JJ Thomson Avenue, Cambridge, CB3 0HE,
United Kingdom}
\affiliation{London Centre for Nanotechnology, Gordon Street, London, WC1H 0AH,
United Kingdom}

\author{Meera M. Parish}
\affiliation{Cavendish Laboratory, JJ Thomson Avenue, Cambridge, CB3 0HE,
United Kingdom}
\affiliation{London Centre for Nanotechnology, Gordon Street, London, WC1H 0AH,
United Kingdom}
\date{\today}

\begin{abstract} 
We consider a two-component gas of fermionic atoms 
confined to a quasi-two-dimensional (quasi-2D) geometry by a harmonic trapping
potential in the transverse direction.  
We construct a mean field theory of the BCS-BEC crossover at zero temperature
that
allows us to
extrapolate to
an infinite number of transverse harmonic oscillator levels.
In the extreme BEC limit, where the confinement length exceeds the dimer size,
we recover 3D dimers confined to
2D with
weak repulsive interactions. 
However, even when the interactions are weak and 
the Fermi energy is less than the confinement frequency,
we find that the higher transverse levels can substantially modify
fermion pairing. 
We argue that recent experiments on pairing in quasi-2D Fermi gases [Y. Zhang et al., Phys.\ Rev.\ Lett.\ \textbf{108}, 235302 (2012)] have already observed the effects of higher transverse levels.
\end{abstract}


\maketitle
\section{Introduction}
\label{sec-introduction}
Recent experimental advances have made it possible to
study 
quasi-two-dimensional (quasi-2D) atomic Fermi gases in a very controlled manner~\cite{Martiyanov2010,DykKWH11,Frohlich2011,2011Natur.480...75F,KosPVF12,BauFFV12,Sommer2011,Zhang2012}. 
Such simple quasi-2D systems may provide useful insights into 
structurally complicated 
unconventional 
superconductors like the cuprates, where superconductivity originates in the copper oxide planes~\cite{Norman2011}.
Quasi-2D Fermi gases are also of fundamental interest, since
they are the marginal case of the Mermin-Wagner theorem and thus have
modified superfluid properties~\cite{PetBS03,OrsS05}. 
In addition, the quasi-2D geometry can strongly affect fermion pairing within the superfluid,
as we investigate in this paper.

In cold-gas experiments, atoms may be confined to one or more quasi-2D ``pancake'' structures using a 
1D 
optical lattice.  
The system can then be tuned from 3D to 2D by increasing 
the confining lattice potential. For sufficiently strong lattices, 
the confining potential for a single quasi-2D layer 
can be modelled as a harmonic oscillator potential, $V(z) = \frac{1}{2} m\omega_z^2z^2$,
in the transverse $z$ direction, where $m$ is the atom mass. 
When the temperature $k_B T \ll \hbar \omega_z$ and the Fermi energy $\varepsilon_F \ll \hbar \omega_z$, the atoms will reside in the lowest harmonic oscillator level in the absence of interactions, and the gas is considered to be kinematically 2D. 
The 
short-range interatomic interactions are also renormalized to yield an effective 2D $s$-wave scattering amplitude and associated two-body bound state with a binding energy $\varepsilon_B$ that depends on both the 3D scattering length and the confinement~\cite{PetS01,BloDZ08}. 
Thus, by varying the ratio $\varepsilon_B/\varepsilon_F$, one can explore the crossover  in 2D from weak BCS pairing 
($\varepsilon_B/\varepsilon_F \ll 1$) to the Bose Einstein condensation (BEC) of dimers ($\varepsilon_B/\varepsilon_F \gg 1$)~\cite{RanDS90,Bertaina2011}.
However, the interactions also mix in higher harmonic levels: for instance, in the BEC limit, dimers will be smaller than the confinement length $l_z = \sqrt{\hbar/m\omega_z}$ 
once $\varepsilon_B > \hbar\omega_z$,  so that they
essentially become 3D bosons confined to quasi-2D. 
Here, we are interested in how the confinement can impact pairing and 
lead to a departure from 
2D behavior throughout the BCS-BEC crossover.

We focus on zero temperature, where there is a well-defined condensate in 2D,
and we construct a mean-field theory that generalizes the 2D results of Randeria \textit{et al.}~\cite{RanDS90} to quasi-2D. 
We expect the mean-field approximation to be reasonable since it 
appears to be consistent with recent experiments in the 2D limit ($\varepsilon_F \ll \hbar\omega_z$)~\cite{Sommer2011}. 
The BCS regime of the quasi-2D Fermi gas was previously studied in Ref.~\cite{MarT05} using a mean-field 
Bogoliubov-de Gennes calculation 
that included the lowest three harmonic oscillator levels. Our approach, however, allows us to extrapolate to an infinite number of levels and thus explore the entire BCS-BEC crossover. 
In the limit $\varepsilon_B \gg \hbar\omega_z$, 
we find that our calculation recovers weakly repulsive 3D bosons confined to quasi-2D. However, 
even for weak interactions, $\varepsilon_B \ll \hbar\omega_z$, we find that higher harmonic levels can substantially modify fermion pairing as we perturb away from pure 2D and $\varepsilon_F$ approaches $\hbar \omega_z$. 
We determine the radio frequency (RF) spectra for the quasi-2D Fermi gas 
and show that recent measurements of pairing in the quasi-2D regime
$\varepsilon_F \sim \hbar\omega_z$~\cite{Zhang2012} are consistent with effects
due to higher transverse levels.

\section{Methodology}
\label{sec-theory}
We consider a two-component ($\uparrow$, $\downarrow$) Fermi
gas interacting close to a broad $s$-wave Feshbach resonance in 3D. Under
harmonic confinment in the $z$-direction, the motion of each atom can be
parameterized by its 2D momentum $\mathbf{k}$ in the $x$-$y$ plane and the
harmonic oscillator quantum number $n$ in the transverse direction.
The many-body Hamiltonian is thus (setting the system volume to 1):
\begin{align}
\label{eq-fullham}
\hat{H}& 
=  \sum_{\mathbf{k}, n, \sigma}(\epsilon_{\mathbf{k} n}-\mu)
c^\dagger_{\mathbf{k} n \sigma} c_{\mathbf{k} n \sigma}\\
& +
\sum_{\substack{\mathbf{k}, n_1, n_2 \\ \mathbf{k}', n_3, n_4 \\ \mathbf{q}}}
\langle n_1 n_2 |\hat{g}| n_3 n_4 \rangle
c^\dagger_{\mathbf{k} n_1 \uparrow}c^\dagger_{\mathbf{q}-\mathbf{k} n_2
\downarrow}
c_{\mathbf{q}-\mathbf{k}' n_3 \downarrow}c_{\mathbf{k}' n_4 \uparrow}\nonumber,
\end{align}
where $\epsilon_{\mathbf{k} n}=k^2/2m + n\omega_z$ 
are the single particle energies relative to the zero-point energy of the $n=0$ state 
(we now set $\hbar=1$), 
and $\mu$ is the chemical potential. 
Note that we assume the mass and chemical potential 
(and thus the particle density) are the same for each spin $\sigma$. 

Since the short-range interactions only depend on the relative motion, we obtain the interaction
matrix elements $\langle n_1 n_2|\hat{g}| n_3 n_4 \rangle$ by
switching to relative and center of mass harmonic oscillator quantum numbers, $\nu$ and $N$
respectively. This yields 
%
\begin{align}
\label{eq-intmatelts}
 \langle n_1 n_2|\hat{g}| n_3 n_4 \rangle
&= g\sum_N f_\nu\langle n_1 n_2|N \nu \rangle
f_{\nu'}\langle N \nu'|n_3 n_4\rangle  \nonumber \\
&\equiv g\sum_N V_N^{n_1 n_2}V_N^{n_3 n_4}, 
\end{align}
where $f_\nu=\sum_{k_z}\tilde{\phi}_\nu(k_z)$, with $\tilde{\phi}_\nu$ 
the Fourier transform of the $\nu$-th harmonic 
oscillator eigenfunction.
It is easily seen that ${f_{2\nu+1}=0}$ and
${f_{2\nu}=\frac{(-1)^\nu}{\nu!}(\frac{m\omega_z}{2\pi})^{1/4}\sqrt{\frac{
(2\nu)! } { 2^ { 2\nu } } } } $. The change of basis coefficients are given in
Ref.~\cite{ChaW67}. Since ${\langle n_1 n_2|N \nu \rangle\sim \delta_{N+\nu,
n_1+n_2}}$ and $\nu,\nu'$ are even, $n_1+n_2$ must equal, modulo 2, $n_3+n_4$  to
obtain a non-zero interaction matrix element. 
The 3D contact interaction $g$ can be written in terms of the binding energy $\eB$ of the 
two-body bound state which always exists 
in the quasi-2D geometry:
\begin{align}
\label{eq-inverseg}
-\frac{1}{g} = \sum_{\textbf{k}, n_1, n_2}\frac{f_{n_1+n_2}^2
|\langle  n_1 n_2 | 0 \hspace*{2mm} n_1+n_2\rangle|^2}{\epsilon_{\textbf{k}
n_1}+\epsilon_{\textbf{k} n_2}+\eB} \>.
\end{align}
Here, we simply take $N=0$ since $\eB$ is independent of the center of mass motion.
One can also determine $\eB$ as a function of the 3D scattering length $a_s$~\cite{PetS01,BloDZ08}
from Eq.~(\ref{eq-inverseg}) using
%
$\frac{1}{g}=\frac{m}{4\pi}\left(\frac{1}{a_s}-\frac{2\Lambda}{\pi} \right)$, 
where $\Lambda$ is a UV cutoff for the 3D momentum that can be sent to infinity
at the end of the calculation.

Now if we define the superfluid order parameter 
\begin{equation}
\Delta_{\mathbf{q} N}=g\sum_{\mathbf{k}, n_1, n_2} V_N^{n_1 n_2} \langle
c_{\mathbf{q}-\mathbf{k}n_2\downarrow}c_{\mathbf{k}n_1\uparrow}\rangle,
\end{equation}
and assume fluctuations around this are small, we obtain the mean-field
Hamiltonian,
\begin{align}
\label{eq-mfham}
\hat{H}_\mathrm{MF} = & \sum_{\mathbf{k}, n, \sigma}(\epsilon_{\mathbf{k}
n}-\mu)
c^\dagger_{\mathbf{k} n \sigma} c_{\mathbf{k} n \sigma}\\ \nonumber
& +
\sum_{\mathbf{q}, N}\bigg(
\Delta_{\mathbf{q} N}\sum_{\mathbf{k}, n_1, n_2}
V_N^{n_1 n_2}c^\dagger_{\mathbf{k} n_1 \uparrow}c^\dagger_{\mathbf{q}-\mathbf{k}
n_2
\downarrow}\\
& +
\Delta_{\mathbf{q} N}^\ast\sum_{\mathbf{k}', n_3, n_4}
V_N^{n_3 n_4}c_{\mathbf{q}-\mathbf{k}'
n_3 \downarrow} c_{\mathbf{k}'n_4 \uparrow}
-\frac{|\Delta_{\mathbf{q} N}|^2}{g}
\bigg).\nonumber
\end{align}
%
We further assume that the ground state has a uniform order parameter without nodes so
that ${\Delta_{\mathbf{q} N}=\delta_{\mathbf{q}
0}  \delta_{N 0}  \Delta_0}$. 
In this case, Eq.~(\ref{eq-mfham}) only contains a single unknown parameter $\Delta_0$, so  
 it can be diagonalized to yield 
\begin{equation}
\label{eq-quasiham}
\hat{H}_\mathrm{MF}=\sum_{\mathbf{k}, n} (\epsilon_{\mathbf{k} n}-\mu-E_{\mathbf{k} n})
-\frac{\Delta_0^2}{g} +
\sum_{\mathbf{k}, n,
\sigma}E_{\textbf{k} n }\gamma^\dagger_{\textbf{k} n \sigma}
\gamma_{\textbf{k} n \sigma}, 
\end{equation}
where 
$E_{\textbf{k} n }$ are the
quasiparticle excitation energies. 
The quasiparticle creation and annihilation operators are respectively 
given by
\begin{align} 
\label{eq-gammaup}
\gamma^\dagger_{\textbf{k} n \uparrow} & =\sum_{n'} (u_{\textbf{k}
n'n}c^\dagger_{\mathbf{k} n' \uparrow}+
v_{\textbf{k} n'n}c_{-\mathbf{k} n' \downarrow}) \\ 
\label{eq-gammadown}
\gamma_{-\textbf{k} n \downarrow} & =\sum_{n'} (u_{\textbf{k} n' n
}c_{-\mathbf{k} n' \downarrow}-
v_{\textbf{k} n'n}c^\dagger_{\mathbf{k} n' \uparrow}) ,
\end{align}
where the real amplitudes $u$, $v$ 
only depend on the magnitude $k \equiv |\vect{k}|$ and satisfy $\sum_{n'} (|u_{\textbf{k}
n'n}|^2 + |v_{\textbf{k} n'n}|^2)=1$. 
Note that while they have a well defined spin and momentum, they involve a superposition of
different harmonic oscillator levels. 
The corresponding BCS wave function $|\Psi_\mathrm{MF}\rangle \propto \prod_{\vect{k}n\sigma} \gamma_{\vect{k}n\sigma} |0\rangle$, where $|0\rangle$ is the vacuum state for the bare operators $c_{\vect{k}n\sigma}$.  
We then minimize 
$\langle \hat{H}_\mathrm{MF} \rangle = \sum_{\mathbf{k}, n} (\epsilon_{\mathbf{k} n}-\mu-E_{\mathbf{k} n}) -\frac{\Delta_0^2}{g}$ with respect to $\Delta_0$ at fixed $\mu$
to obtain the ground state.
The value of $\mu$ is chosen to keep the density of particles 
$\rho=2\sum_{\textbf{k}, n',n} |v_{\textbf{k} n'n}|^2$, 
and thus the Fermi energy $\eF$ 
constant throughout the crossover~\footnote{We define $\eF$ to be the chemical
potential of an ideal Fermi gas with the same density $\rho$.}.

\subsection{Two level calculation}
\label{sec-twoband}
It is first instructive to consider the case of only  two levels $n=0,1$.
Here, Eq.~\eqref{eq-quasiham} is 
greatly simplified since there is no pairing between atoms in the $n=0$
and $n=1$ levels. The quasiparticle dispersions are then 
$E_{\mathbf{k} n}=\sqrt{(\epsilon_{\mathbf{k}n}-\mu)^2+(V^{nn}_0 \Delta_0)^2}$.
One can now minimize $\langle \hat{H}_\mathrm{MF} \rangle$ by simply using 
$\partial \langle \hat{H}_\mathrm{MF} \rangle /\partial\Delta_0  = 0$. 
Combining this with Eq.~\eqref{eq-inverseg} yields the implicit equation
\begin{equation}
\label{eq-del}
\left(\frac{\sqrt{\Delta^2 + \mu^2}-\mu}{\eB} \right)^{4} = 
\frac{\eB
+2\omega_z}{\omega_z-\mu+\sqrt{\frac{\Delta^2}{4} + (\omega_z -
\mu)^2}} 
\> ,
\end{equation}
where we have defined $\Delta=\Delta_0 V_0^{00}$, the pairing gap in the lowest level. 
Also, the density $\rho = -\partial \langle \hat{H}_\mathrm{MF} \rangle /\partial\mu$, and in the regime $\eF\leq \omega_z$ where $\eF = \pi\rho/m$, 
this gives 
%
\begin{equation}
\label{eq-number}
2\eF=\sqrt{\Delta^2 + \mu^2}+\sqrt{\Delta^2/4 + (\omega_z -
\mu)^2}+2\mu-\omega_z.
\end{equation}
%
We see that  Eqs.\ (\ref{eq-del}) and (\ref{eq-number}) reduce to the 2D mean-field equations 
\cite{RanDS90} in the limit $\omega_z \to \infty$, as expected. They also yield the lowest order correction to the 2D result due to confinement ($\eF/\omega_z \neq 0$): in the BCS regime $\eB/\eF \ll 1$, we have: 
%
\begin{align} \notag
\frac{\Delta}{\eF}  \simeq \sqrt{\frac{2\eB}
{\eF}}\left(1+\frac{\eF}{8 \omega_z}\right), \ & \  \
\frac{\mu}{\eF} \simeq 1-\frac{\eB}{2\eF}\left(1+\frac{\eF}{4 \omega_z}
\right).
\end{align}
Thus, 
$\Delta$ is enhanced by the confinement in this regime while $\mu$ is suppressed. 
This trend is also observed in the full calculation involving many levels (see
Figs.~\ref{fig-mu} and \ref{fig-del}).
\subsection{Multiple levels}
\label{sec-multiband}
For more 
accurate results, we must include multiple levels of the confinement, especially when 
perturbing away from the 2D limit $\eF$, $\eB \ll \omega_z$. In general,
$\hat{H}_\mathrm{MF}$ must be diagonalized numerically to obtain 
$E_{\mathbf{k} n}$ and the quasiparticle amplitudes 
for a given $\mu$ and $\Delta$.
Equivalently, one can 
solve the Bogoliubov-de Gennes equations self-consistently, 
but it is considerably faster to minimize the energy $\langle \hat{H}_\mathrm{MF} \rangle$ directly and it also allows us to take into account 
up to 100 levels. 
%
Indeed, we find that higher harmonic levels are important 
even for weak interactions once $\eF/\omega_z$ shifts away from zero.
For the values of $\eB$ and $\eF$ considered in this paper,
we can in fact extrapolate the results for $\mu$ and $\Delta$ to an infinite
number of harmonic levels since we find that they both scale linearly with the
inverse of the number of levels in this limit. We emphasize that the
difference between
extrapolated values and those obtained using only a
few levels can differ by tens of percent, even in the BCS regime (see
Appendix).

\section{Results}
\label{sec-results}
\subsection{Order parameter and chemical potential}
\label{sec-mudel}
\begin{figure}
  \centerline{
  \includegraphics[width=\columnwidth]{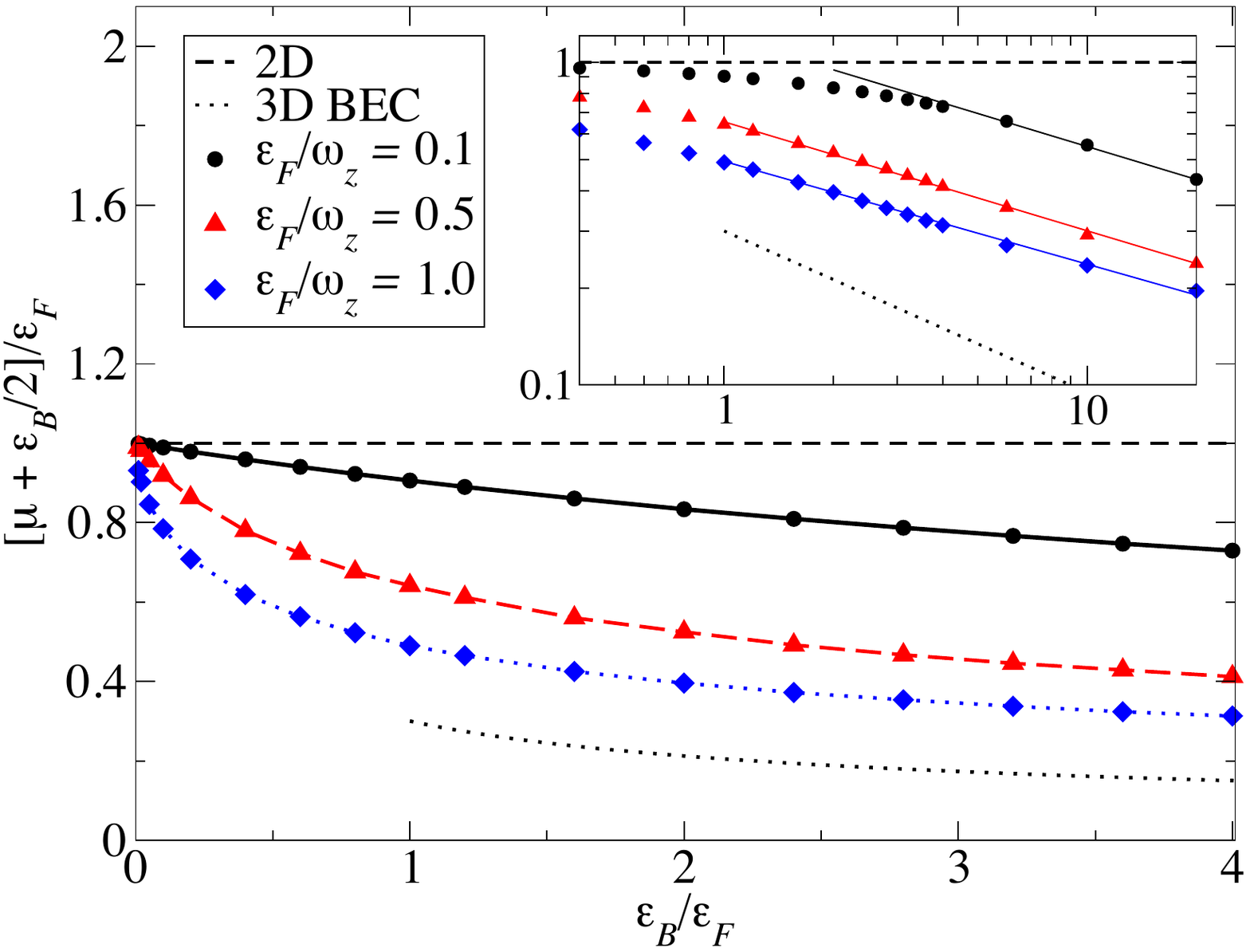} 
  }
  \caption{(Color online) 
Chemical potential $\mu$ measured with respect to half the binding energy $\eB$ 
for several values of the Fermi energy 
$\eF/\omega_z$. The dashed line is the 2D mean-field result 
\cite{RanDS90}, while  
the dotted curve is the 3D BEC result, 
${\mu+\frac{\eB}{2} \simeq \frac{2\sqrt{2}\eF}{3\pi}\sqrt{\frac{\eF}{\eB}}}$. 
Inset: Asymptotic behaviour in the BEC regime $\eB/\eF > 1$ plotted on a
logarithmic scale. Once $\eB/\hbar\omega_z > 1$, we obtain 3D bosons
confined to quasi-2D.
The solid straight lines are straight
line fits to the data with gradient $-1/3$. 
Error bars for the numerical data 
are within symbol size. 
  }
  \label{fig-mu}
\end{figure}

\begin{figure}
  \centerline{
  \includegraphics[width=\columnwidth]{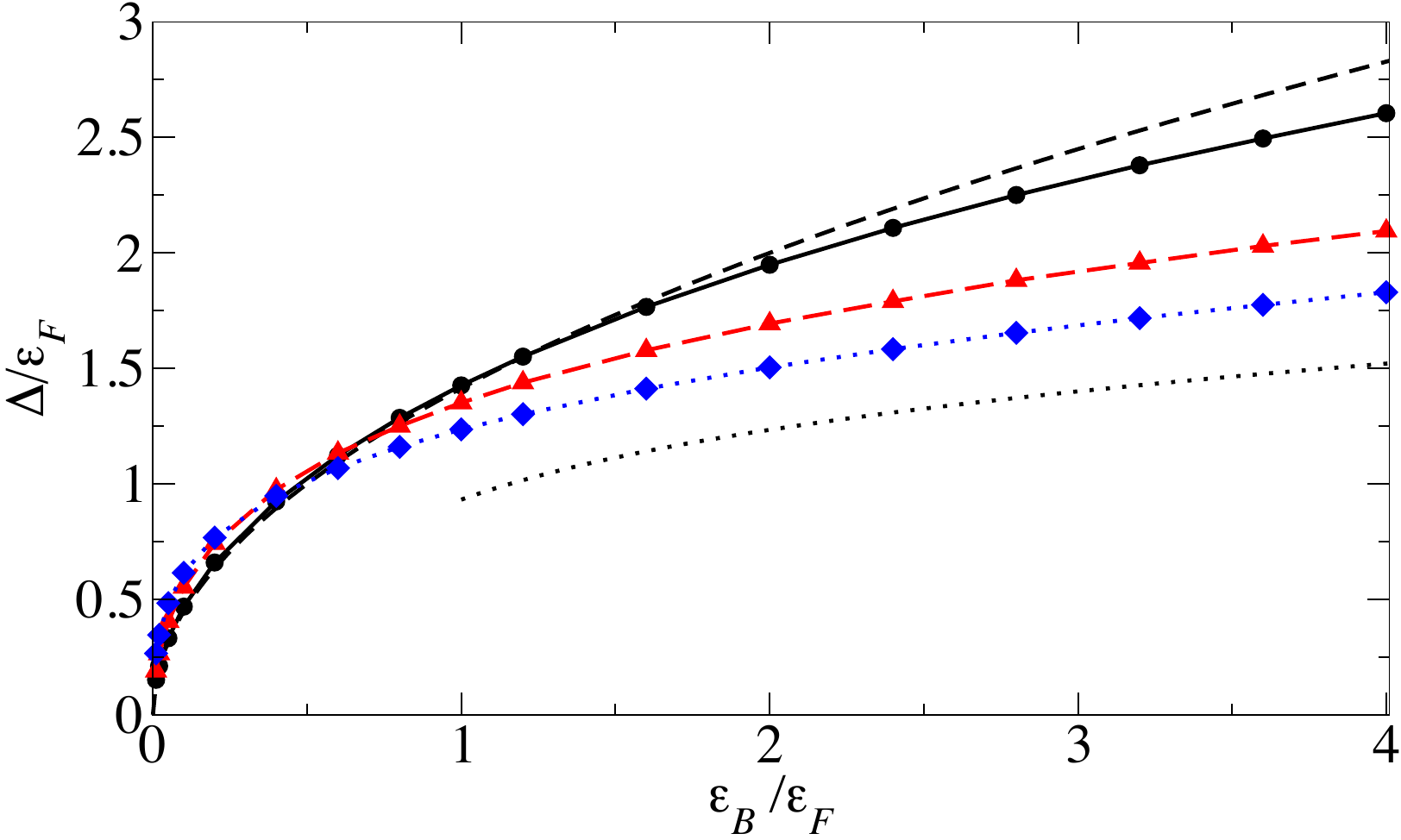} 
  }
  \caption{(Color online) Behavior of the order parameter $\Delta$ throughout the 
BCS-BEC crossover 
for different values of 
$\eF/\omega_z$. The dashed
curve is the 2D mean-field result $\Delta = \sqrt{2\eB\eF}$
\cite{RanDS90}, while 
the dotted curve is the 3D BEC result, $\Delta\simeq \eF\sqrt{\frac{16}{3\pi}} 
\left(\frac{\eB}{2\eF} \right)^\frac{1}{4}$.
The key for the numerical data is the same as in Fig.~\ref{fig-mu}. 
  }
  \label{fig-del}
\end{figure}

By 
incorporating an infinite number of levels, 
we can determine the evolution of the chemical potential $\mu$ and order parameter $\Delta$ throughout the BCS-BEC crossover in a quasi-2D Fermi gas, as depicted in Figs.~\ref{fig-mu} and \ref{fig-del}.
The lowest density we consider ($\eF/\omega_z =0.1$) corresponds approximately to the experiments of Ref.~\cite{Frohlich2011,2011Natur.480...75F,KosPVF12,BauFFV12}. 
Note that in the extreme BCS and BEC limits, we must have $\mu \to \eF$ and $\mu \to -\eB/2$, respectively, a feature which holds across all dimensions.  
In the BCS regime, the behavior is 
in qualitative agreement with the two-level calculation:
$\mu$ is suppressed and $\Delta$ is enhanced with respect to the 2D result, 
with the deviation from 2D being increased with increasing $\eF/\omega_z$.
However, multiple levels are required to correctly capture the dependence on $\eB/\eF$ as $\eF$ 
shifts away from zero. 
We also see in Fig.~\ref{fig-mu} that the relative chemical potential $\mu +\eB/2$ exhibits a very steep gradient as $\eB \to 0$ for $\eF/\omega_z \gtrsim 0.5$.
This illustrates how 
higher levels can lead to a strong deviation from 2D even when $\eF < \omega_z$ and $\eB/\eF \ll 1$. 

For larger $\eB/\eF$, $\Delta$ eventually becomes suppressed compared to the 2D
result and approaches the 3D mean-field curve in the BEC limit
(Fig.~\ref{fig-del}).
The chemical potential, 
on the other hand,
always remains lower than the 2D result, and has a behavior in the BEC regime
that is intermediate between 2D and 3D mean field. 
In the limit $\eB/\eF \gg 1$, 
the pairing gap is no longer given by $\Delta$ and the properties of the Bose superfluid are instead encoded in $\mu$. In particular, the relative quantity $\mu+\eB/2$ yields the mean-field energy for the repulsion between dimers. Referring to Fig.~\ref{fig-mu} (inset), we see that it 
tends to zero as a power law with increasing $\eB/\eF$, similarly to 3D and in contrast to the 2D mean-field result.
This is consistent with a dimensional crossover to 3D dimers once $\eB \gtrsim \omega_z$.
In this case, we expect to have  weakly interacting bosons confined to
quasi-2D with a mean-field 
energy that scales as $a_{s}/l_z \sim \sqrt{\omega_z/\eB}$~\cite{Petrov2000}, i.e., it behaves as a power law with exponent $-1/2$ like in the 3D BEC regime.  
However, our quasi-2D calculation gives a power of $-1/3$ rather than $-1/2$.
%
This discrepancy is due to the fact that our mean-field approximation
does not allow for the scattering of dimers in the transverse direction since
they are constrained to be in the $N=0$ center of mass mode. Removing a
spatial degree of freedom, means we only recover $2/3$ of the full exponent. The
fact
that we recover a power law dependence at all for the
dimer-dimer interaction, is because we can extrapolate to an infinite number of
levels. Note that $N>0$ scattering between dimers is not expected to be
significant in the BCS regime.
The repulsion between dimers 
has also been discussed in the context of a two-channel model for the quasi-2D
system~\cite{Zhang2008}, although no explicit dependence on $\eB$ was given.

\subsection{Experimental Probes}
\label{sec-probes}
\begin{figure}
  \centerline{
  \includegraphics[width=\columnwidth]{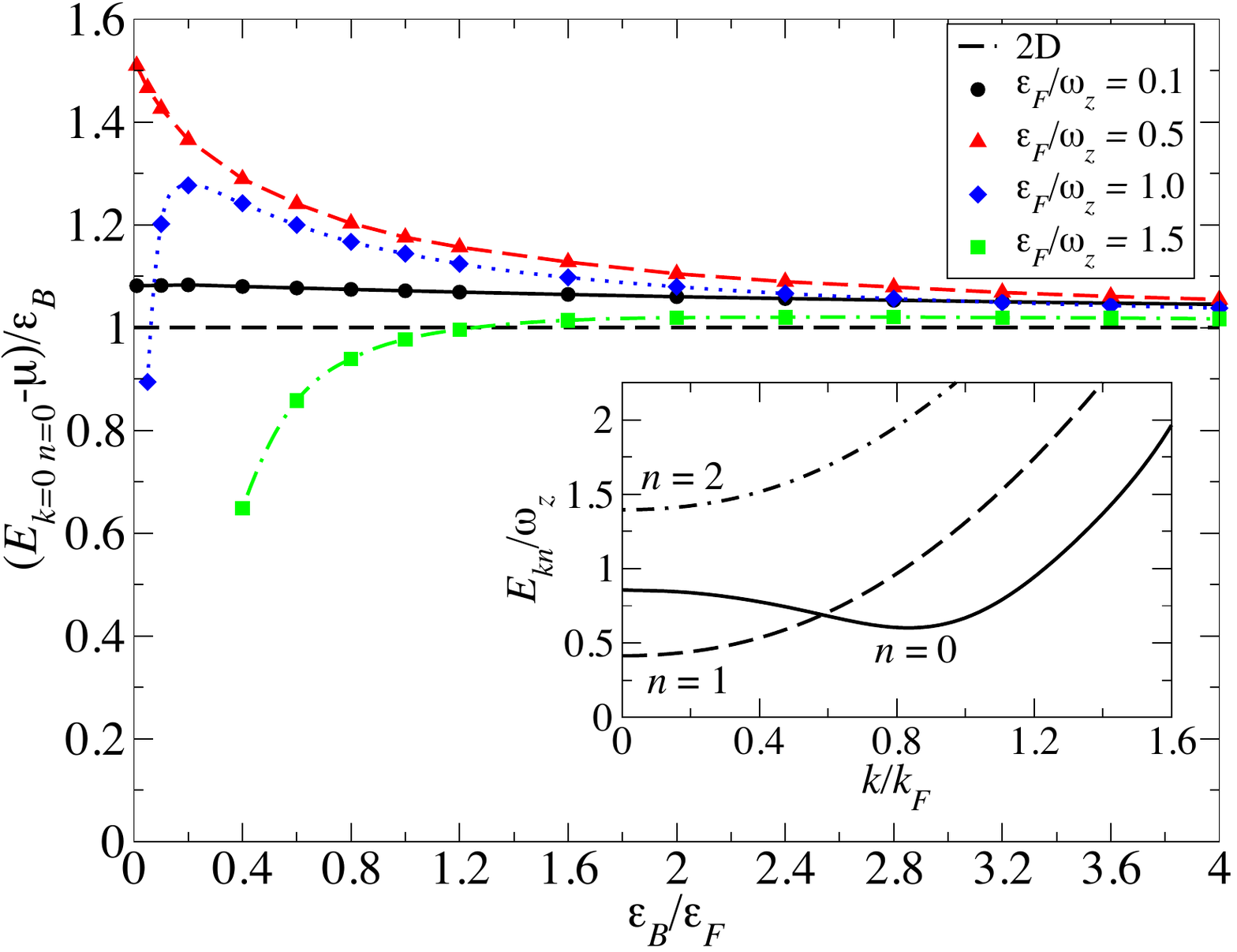} 
  }
  \caption{(Color online) The pairing gap measured using RF spectroscopy as a function of $\eB/\eF$ for a range of different $\eF/\omega_z$. In the BCS regime, it deviates substantially from the 2D result with increasing $\eF/\omega_z$. In the BEC limit, $E_{\mathbf{k}=0,n=0}-\mu$ must always approach $\eB$, regardless of $\eF/\omega_z$.  Inset: Lowest quasiparticle dispersions for $\eF/\omega_z=1$, $\eB/\eF=0.1$. 
Note that there is level repulsion between $n=0$ and $n=2$, but 
no avoided crossing between the $n=0$ and $n=1$ dispersions. 
  }
  \label{fig-exp}
\end{figure}

\begin{figure}
  \centerline{
  \includegraphics[width=\columnwidth]{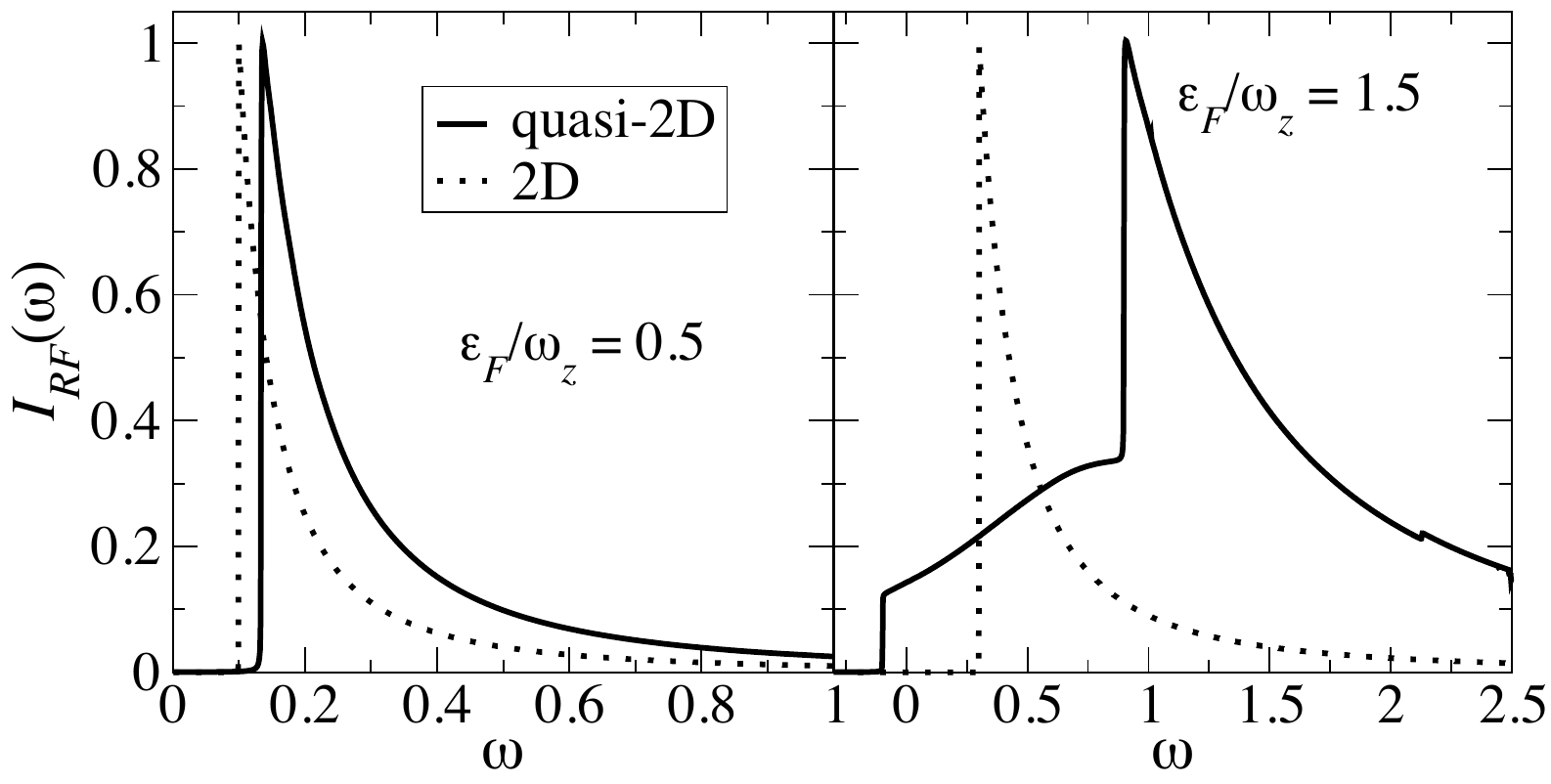} 
  }
  \caption{Radio frequency spectra of the quasi-2D Fermi gas in the BCS regime with $\eB/\eF = 0.2$ for 
$\eF/\omega_z = 0.5,1.5$. 
The RF current $I_{RF}$ is scaled so that the peak value is always 1. 
The peak is strongly shifted towards higher frequencies 
compared to the 2D result when $\eF/\omega_z \simeq 1.5$.
Note that the spectra are calculated assuming that the final state is non-interacting.
  }
  \label{fig-spec}
\end{figure}

Deviations from 2D behavior will also be apparent in experimental probes of the quasi-2D superfluid. Typically, investigations of pairing have exploited RF spectroscopy~\cite{Haussmann2009}, where atoms in one hyperfine spin state (e.g.\ $\downarrow$) are transferred via an RF pulse to another hyperfine state that is initially unoccupied. In the ideal scenario where the final state is non-interacting, 
the mean-field transition rate or RF current is given by
%
\begin{align}\label{eq:RF}
I_{RF}(\omega) & \propto \sum_{\vect{k},n', n} |v_{\vect{k}n'n}|^2 \delta(\epsilon_{\vect{k}n'}-\mu + E_{\vect{k}n}-\omega)
\end{align}
where $\omega$ is the frequency shift relative to the bare transition frequency between hyperfine states. Here, the onset frequency of the RF spectrum corresponds to $E_{\vect{k}=0, n=0} - \mu$ and is 
associated with 
the pairing gap of the superfluid. In the 2D case, \eqref{eq:RF} reduces to 
%
$I_{RF}(\omega)  \propto \frac{\Delta^2}{\omega^2} \Theta(\omega - \varepsilon_B)$
%
and thus the RF pairing gap is simply $\eB$, as noted by Sommer et al.~\cite{Sommer2011}.
Perturbing away from 2D, we find that the RF spectrum can be substantially modifed by the higher confinement levels. 
In Fig.~\ref{fig-exp}, we see that the strongest effects are in the BCS regime, where the pairing gap is initally enhanced compared to $\eB$. Such an enhancement is not surprising given that, in 3D, a pairing gap can exist even when there is no two-body bound state.
However, once $\eF \gtrsim \omega_z$, the pairing gap drops below $\eB$ and even becomes \emph{negative} for small enough $\eB/\eF$. 
This is because the coupling between $n=0$ and $n=2$, 
and the associated level repulsion 
(see inset of Fig.~\ref{fig-exp}), 
reduces the energy $E_{\mathbf{k},n=0}$.
 In this case, the lowest energy quasiparticle contains a smaller fraction of the $n=0$ harmonic level and the RF peak is instead dominated by the $n=2$ quasiparticle.
Thus, the onset frequency is no longer an accurate measure of pairing, as we can see in Fig.~\ref{fig-spec}.
Note that the deep lattices used in Ref.~\cite{Sommer2011} 
correspond to $\eF/\omega_z \approx 0.03$ and thus the pairing gap will lie very close to the 2D result, as was observed.

On the other hand, the experiments in Ref.~\cite{Zhang2012} correspond to $\eF/\omega_z \simeq 1.5$ and thus 
the RF spectrum in the BCS regime will be strongly affected by the confinement. 
In particular, we see in Fig.~\ref{fig-spec} that the RF peak is shifted to higher frequencies compared to the 2D case and develops more structure at lower frequencies. 
Furthermore, the pairing gap in the BEC regime appears to be less sensitive to confinement and closer to the 2D result (Fig.~\ref{fig-exp}) since it is dominated by two-body physics.
These features are all consistent with the experimental observations~\cite{Zhang2012}.
A direct comparison with these experiments is not straightforward in the BCS
regime because of strong final state interactions (which we have neglected). In
addition, thermal effects are expected to smear out the fine structure in the RF
spectra in Fig.~\ref{fig-spec}, leaving an RF peak that more closely resembles
that observed.  
It has even been suggested that finite temperature plays a crucial role here --- see Ref.~\cite{Pie12} for an alternative explanation 
based on fermionic polarons.
However, the shift due to confinement appears to be substantial at $T=0$, with a direction that is consistent with experiment, and therefore it cannot be disregarded. Indeed, 
it has also been shown that effects due to confinement can be significant in 
spin-polarized quasi-2D Fermi gases~\cite{LevB12}.

\begin{figure}
  \centerline{
  \includegraphics[width=0.95\columnwidth]{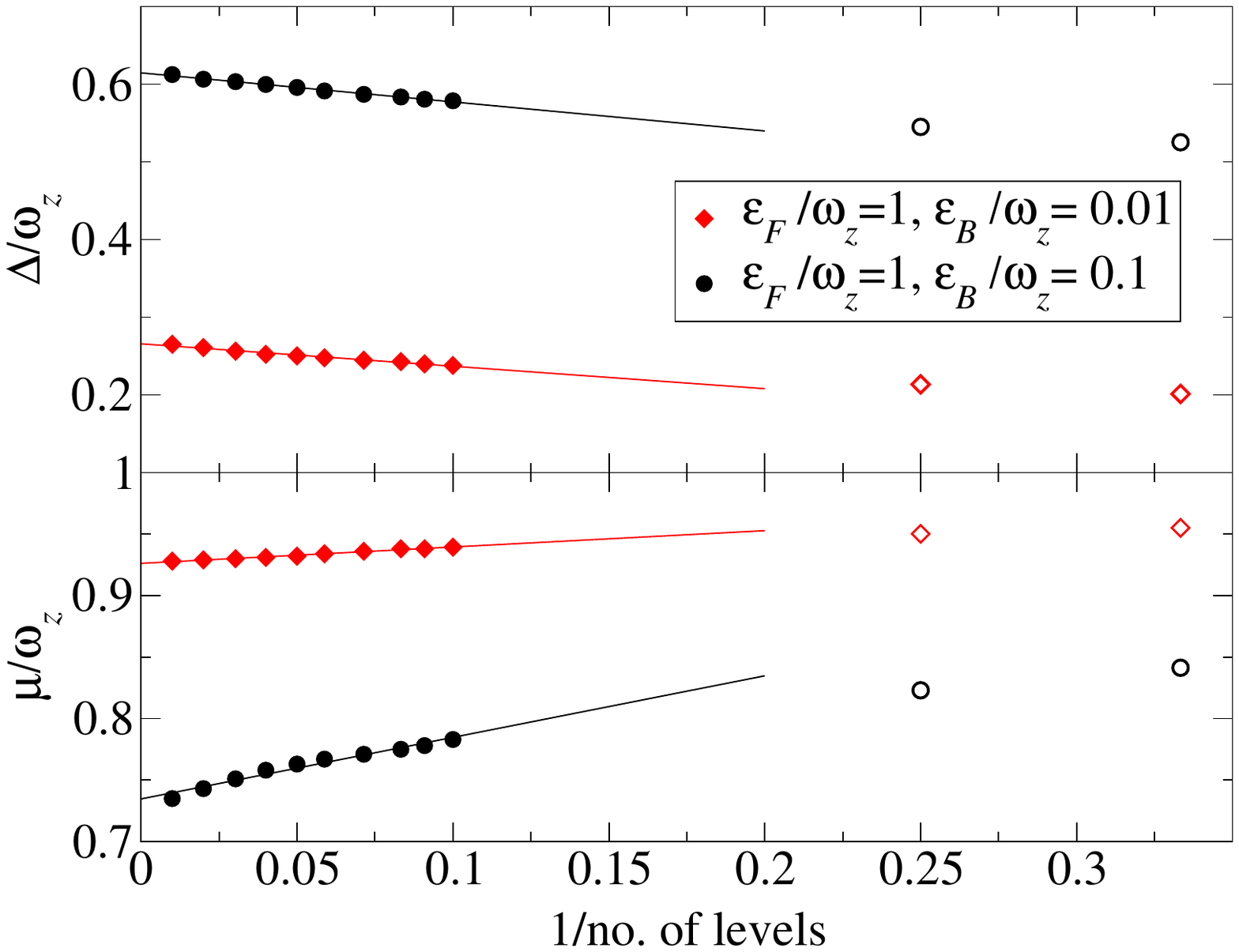} 
  }
  \caption{(Color online) Order parameter, $\Delta$, (upper panel) and chemical
potential, $\mu$, (lower panel) versus inverse number of levels for
$\eF/\omega_z=1$ and $\eB/\omega_z=0.01$ (red diamonds), $\eB/\omega_z=0.1$
(black circles). The solid lines are linear fits to the data for ten or more
levels. The hollow symbols, corresponding to the three and four level cases,
have the same values of $\eB,\eF$ as their solid counterparts, but were not
included in the fit.
  }
  \label{fig-extrapolate}
\end{figure}

\section{Conclusion}
\label{sec-conclusion}
To conclude, we have constructed a mean-field theory for the quasi-2D Fermi gas
that is able to capture the deviations from 2D behavior resulting from
confinement.  
We expect it to provide a benchmark for further investigations into quasi-2D Fermi systems. 
In the future, it would be interesting to explore 
the effects of finite temperature and the Berezinskii-Kosterlitz-Thouless transition in our model.

\begin{acknowledgments}
We gratefully acknowledge fruitful discussions with Michael K\"ohl, Jesper Levinsen, Stefan Baur and Mark Ku.
This work was supported by the EPSRC under Grant No.\ EP/H00369X/2.
\end{acknowledgments}
 
\appendix*

\section{Effect of higher levels}
\label{sec-appendix}
Our approach of directly minimizing the mean field energy, enabled us to do
calculations for up to 100 levels. Indeed, changing the number of levels by one
does not change the results by more than a few
percent, as noted in Ref.~\cite{MarT05}. However, there is an incremental
change if one continues to add more and more levels, resulting in an overall
significant difference. This is to be expected in the BEC regime, but
surprisingly also holds in the BCS regime. Upon plotting $\mu$, $\Delta$ as a
function of the inverse number of levels used in the calculation, we observe a
linear behavior (see Fig.~\ref{fig-extrapolate}). This enables us to
extrapolate to the origin, corresponding to an infinite number of harmonic
levels. For $\eF/\omega_z=1$, as shown in Fig.~\ref{fig-extrapolate}, $\Delta$
changes by $32\%$ for $\eB/\omega_z=0.01$ and $17\%$ for $\eB/\omega_z=0.1$,
when the extrapolated values are compared to those obtained in a three level
calculation. For $\mu$, the differences are $3\%$ for $\eB/\omega_z=0.01$ and
$13\%$ for $\eB/\omega_z=0.1$. These results are for the BCS regime. For the
BEC regime, the percentage changes between a few levels and the extrapolated
values are considerably larger.

\begin{thebibliography}{22}
\expandafter\ifx\csname natexlab\endcsname\relax\def\natexlab#1{#1}\fi
\expandafter\ifx\csname bibnamefont\endcsname\relax
  \def\bibnamefont#1{#1}\fi
\expandafter\ifx\csname bibfnamefont\endcsname\relax
  \def\bibfnamefont#1{#1}\fi
\expandafter\ifx\csname citenamefont\endcsname\relax
  \def\citenamefont#1{#1}\fi
\expandafter\ifx\csname url\endcsname\relax
  \def\url#1{\texttt{#1}}\fi
\expandafter\ifx\csname urlprefix\endcsname\relax\def\urlprefix{URL }\fi
\providecommand{\bibinfo}[2]{#2}
\providecommand{\eprint}[2][]{\url{#2}}

\bibitem[{\citenamefont{Martiyanov et~al.}(2010)\citenamefont{Martiyanov,
  Makhalov, and Turlapov}}]{Martiyanov2010}
\bibinfo{author}{\bibfnamefont{K.}~\bibnamefont{Martiyanov}},
  \bibinfo{author}{\bibfnamefont{V.}~\bibnamefont{Makhalov}}, \bibnamefont{and}
  \bibinfo{author}{\bibfnamefont{A.}~\bibnamefont{Turlapov}},
  \bibinfo{journal}{Phys. Rev. Lett.} \textbf{\bibinfo{volume}{105}},
  \bibinfo{pages}{030404} (\bibinfo{year}{2010}).

\bibitem[{\citenamefont{Dyke et~al.}(2011)\citenamefont{Dyke, Kuhnle, Whitlock,
  Hu, Mark, Hoinka, Lingham, Hannaford, and Vale}}]{DykKWH11}
\bibinfo{author}{\bibfnamefont{P.}~\bibnamefont{Dyke}},
  \bibinfo{author}{\bibfnamefont{E.~D.} \bibnamefont{Kuhnle}},
  \bibinfo{author}{\bibfnamefont{S.}~\bibnamefont{Whitlock}},
  \bibinfo{author}{\bibfnamefont{H.}~\bibnamefont{Hu}},
  \bibinfo{author}{\bibfnamefont{M.}~\bibnamefont{Mark}},
  \bibinfo{author}{\bibfnamefont{S.}~\bibnamefont{Hoinka}},
  \bibinfo{author}{\bibfnamefont{M.}~\bibnamefont{Lingham}},
  \bibinfo{author}{\bibfnamefont{P.}~\bibnamefont{Hannaford}},
  \bibnamefont{and} \bibinfo{author}{\bibfnamefont{C.~J.} \bibnamefont{Vale}},
  \bibinfo{journal}{Phys. Rev. Lett.} \textbf{\bibinfo{volume}{106}},
  \bibinfo{pages}{105304} (\bibinfo{year}{2011}).

\bibitem[{\citenamefont{Fr\"ohlich et~al.}(2011)\citenamefont{Fr\"ohlich, Feld,
  Vogt, Koschorreck, Zwerger, and K\"ohl}}]{Frohlich2011}
\bibinfo{author}{\bibfnamefont{B.}~\bibnamefont{Fr\"ohlich}},
  \bibinfo{author}{\bibfnamefont{M.}~\bibnamefont{Feld}},
  \bibinfo{author}{\bibfnamefont{E.}~\bibnamefont{Vogt}},
  \bibinfo{author}{\bibfnamefont{M.}~\bibnamefont{Koschorreck}},
  \bibinfo{author}{\bibfnamefont{W.}~\bibnamefont{Zwerger}}, \bibnamefont{and}
  \bibinfo{author}{\bibfnamefont{M.}~\bibnamefont{K\"ohl}},
  \bibinfo{journal}{Phys. Rev. Lett.} \textbf{\bibinfo{volume}{106}},
  \bibinfo{pages}{105301} (\bibinfo{year}{2011}).

\bibitem[{\citenamefont{{Feld} et~al.}(2011)\citenamefont{{Feld},
  {Fr{\"o}hlich}, {Vogt}, {Koschorreck}, and {K{\"o}hl}}}]{2011Natur.480...75F}
\bibinfo{author}{\bibfnamefont{M.}~\bibnamefont{{Feld}}},
  \bibinfo{author}{\bibfnamefont{B.}~\bibnamefont{{Fr{\"o}hlich}}},
  \bibinfo{author}{\bibfnamefont{E.}~\bibnamefont{{Vogt}}},
  \bibinfo{author}{\bibfnamefont{M.}~\bibnamefont{{Koschorreck}}},
  \bibnamefont{and}
  \bibinfo{author}{\bibfnamefont{M.}~\bibnamefont{{K{\"o}hl}}},
  \bibinfo{journal}{\nat} \textbf{\bibinfo{volume}{480}}, \bibinfo{pages}{75}
  (\bibinfo{year}{2011}).

\bibitem[{\citenamefont{Koschorreck et~al.}(2012)\citenamefont{Koschorreck,
  Pertot, Vogt, Frohlich, Feld, and Kohl}}]{KosPVF12}
\bibinfo{author}{\bibfnamefont{M.}~\bibnamefont{Koschorreck}},
  \bibinfo{author}{\bibfnamefont{D.}~\bibnamefont{Pertot}},
  \bibinfo{author}{\bibfnamefont{E.}~\bibnamefont{Vogt}},
  \bibinfo{author}{\bibfnamefont{B.}~\bibnamefont{Frohlich}},
  \bibinfo{author}{\bibfnamefont{M.}~\bibnamefont{Feld}}, \bibnamefont{and}
  \bibinfo{author}{\bibfnamefont{M.}~\bibnamefont{Kohl}},
  \bibinfo{journal}{Nature} \textbf{\bibinfo{volume}{485}},
  \bibinfo{pages}{619} (\bibinfo{year}{2012}).

\bibitem[{\citenamefont{Baur et~al.}(2012)\citenamefont{Baur, Fr\"ohlich, Feld,
  Vogt, Pertot, Koschorreck, and K\"ohl}}]{BauFFV12}
\bibinfo{author}{\bibfnamefont{S.~K.} \bibnamefont{Baur}},
  \bibinfo{author}{\bibfnamefont{B.}~\bibnamefont{Fr\"ohlich}},
  \bibinfo{author}{\bibfnamefont{M.}~\bibnamefont{Feld}},
  \bibinfo{author}{\bibfnamefont{E.}~\bibnamefont{Vogt}},
  \bibinfo{author}{\bibfnamefont{D.}~\bibnamefont{Pertot}},
  \bibinfo{author}{\bibfnamefont{M.}~\bibnamefont{Koschorreck}},
  \bibnamefont{and} \bibinfo{author}{\bibfnamefont{M.}~\bibnamefont{K\"ohl}},
  \bibinfo{journal}{Phys. Rev. A} \textbf{\bibinfo{volume}{85}},
  \bibinfo{pages}{061604} (\bibinfo{year}{2012}).

\bibitem[{\citenamefont{Sommer et~al.}(2012)\citenamefont{Sommer, Cheuk, Ku,
  Bakr, and Zwierlein}}]{Sommer2011}
\bibinfo{author}{\bibfnamefont{A.~T.} \bibnamefont{Sommer}},
  \bibinfo{author}{\bibfnamefont{L.~W.} \bibnamefont{Cheuk}},
  \bibinfo{author}{\bibfnamefont{M.~J.~H.} \bibnamefont{Ku}},
  \bibinfo{author}{\bibfnamefont{W.~S.} \bibnamefont{Bakr}}, \bibnamefont{and}
  \bibinfo{author}{\bibfnamefont{M.~W.} \bibnamefont{Zwierlein}},
  \bibinfo{journal}{Phys. Rev. Lett.} \textbf{\bibinfo{volume}{108}},
  \bibinfo{pages}{045302} (\bibinfo{year}{2012}).

\bibitem[{\citenamefont{Zhang et~al.}(2012)\citenamefont{Zhang, Ong, Arakelyan,
  and Thomas}}]{Zhang2012}
\bibinfo{author}{\bibfnamefont{Y.}~\bibnamefont{Zhang}},
  \bibinfo{author}{\bibfnamefont{W.}~\bibnamefont{Ong}},
  \bibinfo{author}{\bibfnamefont{I.}~\bibnamefont{Arakelyan}},
  \bibnamefont{and} \bibinfo{author}{\bibfnamefont{J.~E.}
  \bibnamefont{Thomas}}, \bibinfo{journal}{Phys. Rev. Lett.}
  \textbf{\bibinfo{volume}{108}}, \bibinfo{pages}{235302}
  (\bibinfo{year}{2012}).

\bibitem[{\citenamefont{Norman}(2011)}]{Norman2011}
\bibinfo{author}{\bibfnamefont{M.~R.} \bibnamefont{Norman}},
  \bibinfo{journal}{Science} \textbf{\bibinfo{volume}{332}},
  \bibinfo{pages}{196} (\bibinfo{year}{2011}).

\bibitem[{\citenamefont{Petrov et~al.}(2003)\citenamefont{Petrov, Baranov, and
  Shlyapnikov}}]{PetBS03}
\bibinfo{author}{\bibfnamefont{D.~S.} \bibnamefont{Petrov}},
  \bibinfo{author}{\bibfnamefont{M.~A.} \bibnamefont{Baranov}},
  \bibnamefont{and} \bibinfo{author}{\bibfnamefont{G.~V.}
  \bibnamefont{Shlyapnikov}}, \bibinfo{journal}{Phys. Rev. A}
  \textbf{\bibinfo{volume}{67}}, \bibinfo{pages}{031601}
  (\bibinfo{year}{2003}).

\bibitem[{\citenamefont{Orso and Shlyapnikov}(2005)}]{OrsS05}
\bibinfo{author}{\bibfnamefont{G.}~\bibnamefont{Orso}} \bibnamefont{and}
  \bibinfo{author}{\bibfnamefont{G.~V.} \bibnamefont{Shlyapnikov}},
  \bibinfo{journal}{Phys. Rev. Lett.} \textbf{\bibinfo{volume}{95}},
  \bibinfo{pages}{260402} (\bibinfo{year}{2005}).

\bibitem[{\citenamefont{Petrov and Shlyapnikov}(2001)}]{PetS01}
\bibinfo{author}{\bibfnamefont{D.~S.} \bibnamefont{Petrov}} \bibnamefont{and}
  \bibinfo{author}{\bibfnamefont{G.~V.} \bibnamefont{Shlyapnikov}},
  \bibinfo{journal}{Phys. Rev. A} \textbf{\bibinfo{volume}{64}},
  \bibinfo{pages}{012706} (\bibinfo{year}{2001}).

\bibitem[{\citenamefont{Bloch et~al.}(2008)\citenamefont{Bloch, Dalibard, and
  Zwerger}}]{BloDZ08}
\bibinfo{author}{\bibfnamefont{I.}~\bibnamefont{Bloch}},
  \bibinfo{author}{\bibfnamefont{J.}~\bibnamefont{Dalibard}}, \bibnamefont{and}
  \bibinfo{author}{\bibfnamefont{W.}~\bibnamefont{Zwerger}},
  \bibinfo{journal}{Rev. Mod. Phys.} \textbf{\bibinfo{volume}{80}},
  \bibinfo{pages}{885} (\bibinfo{year}{2008}).

\bibitem[{\citenamefont{Randeria et~al.}(1990)\citenamefont{Randeria, Duan, and
  Shieh}}]{RanDS90}
\bibinfo{author}{\bibfnamefont{M.}~\bibnamefont{Randeria}},
  \bibinfo{author}{\bibfnamefont{J.-M.} \bibnamefont{Duan}}, \bibnamefont{and}
  \bibinfo{author}{\bibfnamefont{L.-Y.} \bibnamefont{Shieh}},
  \bibinfo{journal}{Phys. Rev. Lett.} \textbf{\bibinfo{volume}{62}},
  \bibinfo{pages}{981} (\bibinfo{year}{1989});
  \bibinfo{journal}{Phys. Rev. B} \textbf{\bibinfo{volume}{41}},
  \bibinfo{pages}{327} (\bibinfo{year}{1990}).

\bibitem[{\citenamefont{Bertaina and Giorgini}(2011)}]{Bertaina2011}
\bibinfo{author}{\bibfnamefont{G.}~\bibnamefont{Bertaina}} \bibnamefont{and}
  \bibinfo{author}{\bibfnamefont{S.}~\bibnamefont{Giorgini}},
  \bibinfo{journal}{Phys. Rev. Lett.} \textbf{\bibinfo{volume}{106}},
  \bibinfo{pages}{110403} (\bibinfo{year}{2011}).

\bibitem[{\citenamefont{Martikainen and T\"orm\"a}(2005)}]{MarT05}
\bibinfo{author}{\bibfnamefont{J.-P.} \bibnamefont{Martikainen}}
  \bibnamefont{and}
  \bibinfo{author}{\bibfnamefont{P.}~\bibnamefont{T\"orm\"a}},
  \bibinfo{journal}{Phys. Rev. Lett.} \textbf{\bibinfo{volume}{95}},
  \bibinfo{pages}{170407} (\bibinfo{year}{2005}).

\bibitem[{\citenamefont{Chasman and Wahlborn}(1967)}]{ChaW67}
\bibinfo{author}{\bibfnamefont{R.}~\bibnamefont{Chasman}} \bibnamefont{and}
  \bibinfo{author}{\bibfnamefont{S.}~\bibnamefont{Wahlborn}},
  \bibinfo{journal}{Nuclear Physics A} \textbf{\bibinfo{volume}{90}},
  \bibinfo{pages}{401 } (\bibinfo{year}{1967}), ISSN \bibinfo{issn}{0375-9474}.

\bibitem[{\citenamefont{Petrov et~al.}(2000)\citenamefont{Petrov, Holzmann, and
  Shlyapnikov}}]{Petrov2000}
\bibinfo{author}{\bibfnamefont{D.~S.} \bibnamefont{Petrov}},
  \bibinfo{author}{\bibfnamefont{M.}~\bibnamefont{Holzmann}}, \bibnamefont{and}
  \bibinfo{author}{\bibfnamefont{G.~V.} \bibnamefont{Shlyapnikov}},
  \bibinfo{journal}{Phys. Rev. Lett.} \textbf{\bibinfo{volume}{84}},
  \bibinfo{pages}{2551} (\bibinfo{year}{2000}).

\bibitem[{\citenamefont{Zhang et~al.}(2008)\citenamefont{Zhang, Lin, and
  Duan}}]{Zhang2008}
\bibinfo{author}{\bibfnamefont{W.}~\bibnamefont{Zhang}},
  \bibinfo{author}{\bibfnamefont{G.-D.} \bibnamefont{Lin}}, \bibnamefont{and}
  \bibinfo{author}{\bibfnamefont{L.-M.} \bibnamefont{Duan}},
  \bibinfo{journal}{Phys. Rev. A} \textbf{\bibinfo{volume}{77}},
  \bibinfo{pages}{063613} (\bibinfo{year}{2008}).

\bibitem[{\citenamefont{Haussmann et~al.}(2009)\citenamefont{Haussmann, Punk,
  and Zwerger}}]{Haussmann2009}
\bibinfo{author}{\bibfnamefont{R.}~\bibnamefont{Haussmann}},
  \bibinfo{author}{\bibfnamefont{M.}~\bibnamefont{Punk}}, \bibnamefont{and}
  \bibinfo{author}{\bibfnamefont{W.}~\bibnamefont{Zwerger}},
  \bibinfo{journal}{Phys. Rev. A} \textbf{\bibinfo{volume}{80}},
  \bibinfo{pages}{063612} (\bibinfo{year}{2009}).

\bibitem[{\citenamefont{Pietil\"a}(2012)}]{Pie12}
\bibinfo{author}{\bibfnamefont{V.}~\bibnamefont{Pietil\"a}},
  \bibinfo{journal}{Phys. Rev. A} \textbf{\bibinfo{volume}{86}},
  \bibinfo{pages}{023608} (\bibinfo{year}{2012}).

\bibitem[{\citenamefont{Levinsen and Baur}(2012)}]{LevB12}
\bibinfo{author}{\bibfnamefont{J.}~\bibnamefont{Levinsen}} \bibnamefont{and}
  \bibinfo{author}{\bibfnamefont{S.~K.} \bibnamefont{Baur}},
  \bibinfo{journal}{Phys. Rev. A} \textbf{\bibinfo{volume}{86}},
  \bibinfo{pages}{041602} (\bibinfo{year}{2012}).

\end{thebibliography}

\end{document}